\documentclass[journal,12pt,onecolumn,draftclsnofoot]{IEEEtran}
\usepackage{graphics}
\usepackage{graphicx}
\usepackage{subcaption}
\usepackage{epsfig}
\usepackage{lipsum,amsmath,multicol}
\usepackage{dblfloatfix}
\usepackage{epstopdf}
\usepackage{mathrsfs}
\usepackage{amsfonts, amssymb}
\usepackage[displaymath, mathlines,running]{lineno}
\usepackage{amsmath}
\usepackage{mathtools}
\usepackage{mathtools}
\usepackage{multirow}
\usepackage[sort]{cite}

\DeclarePairedDelimiter{\floor}{\lfloor}{\rfloor}
\DeclareMathOperator{\E}{\mathbb{E}}

\makeatletter
\newcommand*{\rom}[1]{\expandafter\@slowromancap\romannumeral #1@}
\makeatother

\begin{document}

\title{Throughput of CDM-based Random Access With SINR Capture}

\author{Hoesang~Choi and Hichan~Moon,~\IEEEmembership{Member,~IEEE}
\thanks{Hoesang Choi and Hichan Moon are with the Department of Electronic Engineering, Hanyang University, Seoul, South Korea (e-mail: sanchoi@hanyang.ac.kr and hcmoon@hanyang.ac.kr).}
}

\IEEEcompsoctitleabstractindextext{
\begin{abstract}
\end{abstract}
\end{IEEEkeywords}}

\maketitle
\begin{abstract}
Code division multiplexing (CDM)-based random access is used in many practical wireless systems.
With CDM-based random access, a set of sequences is reserved for random access. A remote station transmits a random access packet using a randomly selected sequence among the set. If more than one remote stations transmit random access packets using the same sequence simultaneously, performance degrades due to sequence collision. In addition, if more than one remote stations transmit random access packets using different sequences simultaneously, performance also degrades due to interference. Therefore, the performance of CDM-based random access is dependent on both sequence collision and interference. There has been no previous research to analyze the performance of CDM-based random access considering both sequence collision and interference. In this paper, throughput of CDM-based random access is investigated considering both sequence collision and interference based on a signal to interference plus noise ratio (SINR) capture model. Analysis and numerical simulation compare the throughputs of several random access schemes including conventional and channel-adaptive random access. The results show that channel-adaptive random access can achieve significantly higher throughput than conventional random access. In addition, based on the results of this paper, it is possible to analyze the trade-off between the throughput and implementation complexity with increased number of sequences.
\end{abstract}

\IEEEpeerreviewmaketitle

\section{Introduction}

In a wireless system, there are two different kinds of channels to transmit data. One is a dedicated channel and the other is a random access channel. It is more efficient to use a dedicated channel to transmit a large amount of data or frequently generated message. On the other hand, it is more efficient to use a random access channel to transmit infrequently generated short message.

As many Internet of Things (IoT) services have recently emerged in wireless systems, the number of remote stations is dramatically increasing within a fixed area of a wireless system \cite{IoT1, 50bil, preamble}. One of the essential requirements of the 5G system is massive connectivity \cite{IoT, massive con}.

Furthermore, the amount of infrequently generated short message is continuously increasing \cite{random access short message2}. For example, a short packet is usually used in many IoT services including power metering, weather monitoring and health care services \cite{rach application ref}. Since random access is efficient for infrequently generated short message, it is being more widely used in many wireless systems. Because of this, it is essential to increase the throughput of random access, which is the number of successful random access transmissions per unit time.

As the importance of random access has been increasing, there have been many researches to enhance random access. Especially, to reduce power consumption and increase coverage of random access, channel-adaptive random access was proposed in \cite{CARA}. With channel-adaptive random access, a remote station transmits a random access packet only when channel gain is larger than or equal to a predetermined threshold. Otherwise, even though a triggering event occurs, a remote station delays transmission of a random access packet until channel gain becomes larger than or equal to the threshold. Several researches on channel-adaptive random access have been performed \cite{CARA, CARA2, CARA5}. However, the throughput of channel-adaptive random access has not been seriously studied before.



It is essential to evaluate the throughput of random access accurately. There have been many researches on the throughput of random access. In \cite{ALOHA}, pure ALOHA was introduced for a packet switching network. To increase the throughput of pure ALOHA, slotted ALOHA was proposed in \cite{S-ALOHA}. With slotted ALOHA, a remote station transmits a packet only at the beginning of a time slot. In \cite{ALOHA throughput}, throughput and delay were investigated for pure or slotted ALOHA.
The throughput of narrow-band multi-channel slotted ALOHA was investigated in \cite{multichannel random access, multichannel aloha random access, multichannel aloha random access1}. With multi-channel slotted ALOHA, a remote station randomly selects a channel among many available channels for a random access transmission.
To achieve higher system capacity, a cellular system is designed based on frequency reuse, with which the same frequency is used again in some neighboring cells. In \cite{multichannel random access}, throughput of random access was investigated considering both frequency reuse and multi-channel slotted ALOHA. In all the previous researches \cite{ALOHA,S-ALOHA,ALOHA throughput,multichannel random access, multichannel aloha random access, multichannel aloha random access1}, if multiple remote stations transmit random access packets simultaneously using the same channel, it is assumed that none of random access packets are successfully detected because of collision.

To mitigate performance degradation due to collision, multiple sequences are used for random access \cite{CDMA random access, MAI1, MAI2, cdm-based}. A remote station transmits a random access packet with the sequence, which is randomly selected among the available sequences. Therefore, if remote stations simultaneously transmit random access packets with distinct sequences, collision does not occur. However, if more than one remote stations select the same sequence, sequence collision occurs. Since more than one packets are simultaneously transmitted with multiple sequences in the same resource, this random access scheme can be named as CDM-based random access \cite{cdm-based}.

In a cellular system, most resources are managed and allocated by a base station. Therefore, it is easy to predict the interference level at the resource allocated to a remote station, if its location is fixed. However, the resource allocated for random access is shared by many remote stations. For example, the resource can be shared by more than ten thousand remote stations. In the resource, since remote stations independently transmit random access packets, the interference level is a random variable depending on the number of simultaneous random access attempts, which is difficult to control at a base station \cite{LTE book}. Furthermore, at a base station, detection performance degrades due to interference from the correlation between different sequences used for random access \cite{cdm-based}.

  Let us consider random access of a Long-Term Evolution (LTE) system. To reduce the interference, orthogonal sequences are used in an LTE system \cite{LTE random access format}. The orthogonal sequences are generated from root Zadoff-Chu sequences with cyclic shifts \cite{zadoff-chu1,zadoff-chu}. However, orthogonality between random access sequences is not maintained at a base station receiver, since the uplink of a cellular system is not synchronized at the moment of random access transmission \cite{cdm-based}.
In a frequency selective channel, there is also no way to maintain orthogonality between the sequences \cite{LTE book, MAI1,zadoff-chu1}. In addition, since there can exist large amount of frequency offset due to Doppler spread, orthogonality between the sequences is not guaranteed in many practical environments \cite{zadoff-chu,loss of orthogonality, freq. offset,zadoff orthogonal}. Furthermore, since the amount of the cyclic shift between random access sequences is determined by cell radius, Zadoff-Chu sequences generated from multiple root numbers should be used for a random access in a large cell, where Zadoff-Chu sequences from different roots are not orthogonal \cite{LTE random access format}.  Therefore, in a practical system, orthogonal sequences only slightly improve the performance and it is more reasonable to assume interference between different sequences even when the orthogonal sequences are used for random access \cite{cdm-based}. Then, even in an LTE system, throughput of random access should be investigated considering the interference resulting from the correlation between different sequences.

Throughput was investigated for CDM-based random access considering interference \cite{CDMA random access, Bit-to-bit,random access Interference1, random access Interference2,RA comparison1, RA comparison2}. With CDM-based random access, a remote station selects a channel (or sequence) among many available channels (or sequences) in a system.
 The probability of bit error was analyzed using improved-Gaussian approximation \cite{Bit-to-bit}. Based on probabilistic analysis, throughput was investigated for CDM-based random access \cite{CDMA random access, random access Interference1, random access Interference2}. Performance are compared between CDM-based random access and narrow-band multi-channel slotted ALOHA \cite{RA comparison1, RA comparison2}. In these papers, it is assumed that the probability of collision is negligible and that the performance is mainly dependent on interference generated from other remote stations \cite{Bit-to-bit,random access Interference1, random access Interference2,RA comparison1, RA comparison2}. Sequence collision has not been considered in these papers.

  In the previous papers, either sequence collision or interference is considered for random access. However, as it is mentioned above, the performance of random access is dependent on both sequence collision and interference in most cellular systems.

In this paper, throughput of random access is investigated considering sequence collision as well as interference. When there is no sequence collision, an SINR capture model\footnote{In this paper, the performance of a random access packet is analyzed based on an SINR capture model, of which performance is slightly different from that of a real model. It is desirable to consider a more realistic model. However, its performance is highly dependent on the structure of a random access packet, since detection performance should be considered for random access. For example, when a random access packet consists of only preamble, the detection performance of random access is expressed with a generalized Marcum-Q function  \cite{CARA}. In addition, when a random access packet consists of both preamble and message , the detection performance of random access is expressed with a very complex equation based on a generalized Marcum-Q and a Q function \cite{preamble message}. In this paper, an SINR capture model is considered, since this is the first approach to consider both collision and interference, and the SINR capture model is enough to give insight on throughput considering both collision and interference. In \cite{preparing}, we are preparing another paper analyzing the throughput of random access based on more realistic performance models including \cite{CARA, preamble message}.}
 is used to analyze the performance of random access. With the SINR capture model, it is assumed that a random access packet is successfully detected at a host station if received SINR is larger than a required SINR. In \cite{SINR capture1,SINR capture2,SINR capture3,SINR capture4}, an SINR capture model was used to analyze the performance of packet detection. These papers address the validity of the SINR capture model for packet detection. Therefore, in this paper, it is assumed that a random access packet is successfully detected only when the received SINR is larger than a required SINR and sequence collision does not occur.

The contributions of this paper are as follows:
\begin{itemize}
  \item A new system model is presented for random access. This is the first reasonable model in random access research considering both collision and multiple access interference.
  \item The probability of successful packet detection is derived for conventional and channel-adaptive random access using the SINR capture model. Based on analyses on the probability, throughputs of several random access schemes are computed considering both sequence collision and interference. The throughput results are more realistic than those obtained using conventional models.
  \item Throughput of channel-adaptive random access is computed with constant and channel inversion power allocation. It is shown that channel-adaptive random access can achieve significantly higher throughput than conventional random access. Especially, with channel-adaptive random access using channel inversion power allocation, about 1.85 times higher throughput can be achieved compared with conventional random access.
  \item
      With CDM-based random access, as the number of sequences increases, the throughput increases due to reduced collision. However, the complexity of a base station increases with the increased number of sequences. There has been no previous researches investigating the trade-off between the implementation complexity and throughput with multiple random access sequences. In this paper, it is observed that throughput is limited by not only the number of sequences, but also interference. Based on the analysis of this paper, it is possible to estimate a reasonable number of random access sequences considering the trade-off between implementation complexity and throughput.

\end{itemize}

The remainder of this paper is organized as follows. Section \rom{2} presents the system model used in this paper. Section \rom{3} analyzes the throughput of conventional and channel-adaptive random access. Section \rom{4} presents numerical results. Finally, Section \rom{5} concludes this paper.

\section{System Model}
Consider a time division duplex (TDD) cellular system consisting of a host station and $M$ remote stations. The host station periodically transmits a pilot signal. Using the pilot signal, a remote station synchronizes its timing to that of the host station \cite{RA procedure}. A remote station measures the small-scale channel gain and path loss of a downlink. From the measured downlink channel gain, it is possible for a remote station to estimate uplink channel gain from channel reciprocity \cite{reciprocity1, reciprocity2}.

Each remote station independently transmits a random access packet to a host station when a triggering event occurs. Therefore, among $M$ remote stations, multiple remote stations can transmit random access packets simultaneously. Let us denote by $K$ the number of simultaneous random access transmissions among $M$ remote stations. Then, $K$ can be approximated by a random number following a Poisson distribution \cite{multichannel random access, poisson}. If an average packet arrival rate is $\lambda$ per one random access slot, the probability density function (pdf) of $K$ becomes\footnote{When $M$ is much larger than $\lambda$, which is general in most practical systems, $\sum_{k=0}^{M}f_K(k|\lambda)$ can be approximated by 1, since $f_K(k|\lambda)$ is negligible for $k>M$.}
\begin{equation}
f_K(k|\lambda) =
  \begin{cases}
   \frac{\lambda^ke^{-\lambda}}{k!}, & \text{for} \, k = 0,1,2,\cdots,M, \\
   0, &$otherwise$.
  \end{cases}
\end{equation}

A remote station randomly selects a sequence among $N_{\textrm{seq}}$ sequences for random access transmission. When there are more than two remote stations transmitting random access packets in the same random access slot simultaneously, performance degrades. Two different scenarios are possible for the performance degradation. One is the case when the random access packets are transmitted using the same sequence, which results in sequence collision. The other is the case when random access packets are transmitted using different sequences. Then, there is no sequence collision, but performance degrades due to the interference from the packets of the other remote stations.

When there are $K$ remote stations transmitting random access packets in a random access slot simultaneously, the sequence collision probability $p_{\textrm{coll}}(N_{\textrm{seq}},K)$ is computed as\cite{collision3}
\begin{equation}
p_{\textrm{coll}}(N_{\textrm{seq}},K)=1- \left( 1-\frac{1}{N_{\textrm{seq}}}\right)^{K-1}.
  \end{equation}

 Each random access packet is transmitted over a slow time-selective fading channel. Therefore, channel gain is assumed to be constant during a random access packet. In addition, all random access packets transmitted from different remote stations experience independent fading channels. The channel gain $y_i$ for the packet from the $i$-th remote station is expressed as
 \begin{equation}
y_i = \frac{1}{L_i}g_i, \, i=0,1,2,\dots M-1,
 \end{equation}
 where $L_i$ and $g_i$ are its path loss and small-scale channel gain, respectively. The small-scale channel gain is normalized to satisfy $\E[g_i\cdot g_i^*]=1$.  In this paper, it is assumed that each random access packet experiences an independent Rayleigh fading channel. Then, the pdf of the small-scale channel gain is $f_G(g_i)=\textrm{exp}(-g_i)$.

With channel-adaptive random access, a remote station transmits a random access packet only when channel gain is larger than a predetermined transmission threshold $g_{\text{th}}$ \cite{CARA}. In this paper, constant and channel inversion power allocations are considered for channel-adaptive random access \cite{CARA}. When $g_{\text{th}}$ is set to zero, channel-adaptive random access using constant power allocation is exactly the same as conventional random access.

Each remote station transmits a random access packet of length $T_\text{p}$ to a host station. A random access packet includes $N_\text{b}$ information bits \cite{multichannel random access, random access short message2}. If a random access packet is composed of only a preamble as in an LTE system \cite{LTE random access format}, then $N_\text{b}$ is defined as 1. Processing gain $N$ is defined as $\frac{B_\text{w}T_\text{p}}{N_\text{b}}$, where $B_\text{w}$ is the bandwidth of a random access packet.

Let us consider that $K$ remote stations transmit random access packets among $M$ remote stations. Without loss of generality, it can be assumed that the first $K$ remote stations ($i = 0,1,2, \cdots ,K-1$) transmit random access packets.

The sequence used by the $i$-th remote station is denoted by $c_i(t)$, which is selected among $N_{\text{seq}}$ sequences.
Each information bit $d_{i,k}$ of the $i$-th remote station is multiplied by the sequence $c_i(t-kT_{\text{b}})$ ($k=0,1,2, \cdots, N_\text{b}-1$), where $T_{\text{b}}$ is the duration of a bit ($T_{\text{b}}=T_{\text{p}}/N_{\text{b}}$). Then, the transmitted random access packet is expressed as $\sqrt{L_iP(g_i)}\cdot s_i(t)$, where $P(g_i)$ is the allocated transmission power for small-scale channel gain $g_i$ and $s_i(t)=\sum^{N_{\text{B}}-1}_{k=0}d_{i,k}\cdot c_i(t-kT_{\text{b}})$, which is normalized to unit power $\left( \int^{T_\text{p}}_{0}|s_i(t)|^2dt/T_\text{p}=1\right)$. Then, the energy consumed to transmit a random access packet is $L_iP(g_i)T_\text{p}$.

 Therefore, the received signal at a host station can be expressed as
  \begin{equation}
   r(t) =\sum^{K-1}_{j=0}\sqrt{g_jP(g_j)}e^{i\theta_{j}}\cdot s_j(t)+ n(t),
  \end{equation}
    where $n(t)$ is an additive white Gaussian noise with power spectral density $N_0$. Here, $\theta_j$ is the phase component of the $j$-th channel and uniformly distributed in $[0,2\pi)$. The received signal $r(t)$ is independent of path loss $L_i$, since the transmission power for the $i$-th remote station is $L_iP(g_i)$.

 \begin{figure}[t]
  \centering
     \epsfig{file=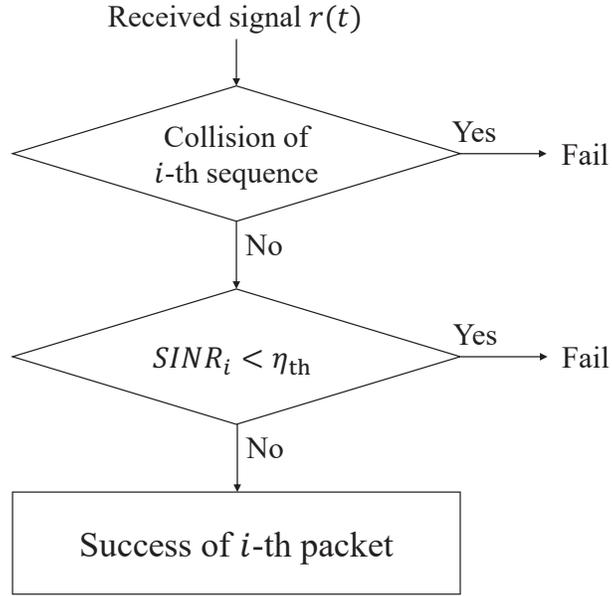, trim=0.0cm 0.0cm 0.0cm 0.0cm, clip=true, width=9cm}
	\caption{Random access detection with SINR capture model.}
\end{figure}

 In this paper, an SINR capture model \cite{SINR capture1,SINR capture2,SINR capture3,SINR capture4} is used as in Fig. 1. With this model, it is assumed that a random access packet is successfully detected at a host station if received SINR is larger than a predetermined SINR threshold $\eta_{\text{th}}$. If the received SINR is smaller than $\eta_{\text{th}}$, it is assumed that a random access packet is not detected successfully at a host station. Therefore, a random access packet can be detected successfully at a host station, if there is no sequence collision and its received SINR is larger than $\eta_{\text{th}}$.

\section{Throughput Analysis}
In this section, throughput is analyzed for conventional and channel adaptive random access schemes.  Throughput is defined as the average number of successful packet transmissions for a random access slot of length $T_\text{p}$.

Since all detector outputs are independent and identically distributed random variables at a host station, without loss of generality, it is possible to consider the performance of the $0$-th detector as in \cite{SINR capture4}. The received signal is despreaded by the assigned sequence $c_0(t)$. Then, for the $0$-th signal detection, the despreaded output $z_0$ can be expressed as
\begin{equation}
z_0=\sqrt{g_0P_0(g_0)}+\sum^{K-1}_{j=1}\sqrt{g_jP_j(g_j)}X_{j0}+n'(t),
\end{equation}
  where $n'(t)=\frac{1}{T_\text{b}}\int^{T_\text{b}}_{0}n(t)c^*_0(t)dt$ and $X_{j0}=\frac{1}{T_\text{b}} \int^{T_\text{b}}_{0}c_j(t)c^*_0(t)e^{i(\theta_j-\theta_0)}dt$. If $N$ is large, which is general in most practical systems, the $X_{j0}$ can be approximated by an independent zero-mean Gaussian random variable with variance $1/N$ when $c_j(t)\neq c_0(t)$ \cite{random access Interference1, accurate ds-cdma}.
  Therefore, the received SINR of $z_0$ is computed as
\begin{equation}
 SINR_0 = \frac{g_0P_0(g_0)}{\displaystyle\frac{N_0}{T_\text{p}} + \frac{1}{N}\sum^{K-1}_{j=1}g_jP_j(g_j)}.
 \end{equation}

 With an SINR capture model, if $K$ simultaneous packets are transmitted with distinct sequences, the probability of successful detection $p_s(K)$ is expressed for each random access packet as
\begin{equation}
 p_s(K) =\textrm{Pr}\{ SINR_0>\eta_{\text{th}}|K, \text{no collision} \}.
  \end{equation}

Throughput $S(\lambda)$ can be expressed as for an average arrival rate $\lambda$ \cite{RA comparison1}
\begin{align}
S(\lambda)&=\E_K[ K\cdot \textrm{Pr}\{SINR_0>\eta_{\text{th}} \, ,\, \text{no collision}|K\}|\lambda]\nonumber \\
&=\E_K[ K\cdot \textrm{Pr}\{ SINR_0>\eta_{\text{th}}|K, \text{no collision} \}\cdot\textrm{Pr}\{\text{no collision}|K\}|\lambda]\nonumber\\
&=\E_K[K\cdot p_s(K)\cdot \left(1-p_{\textrm{coll}}(N_{\textrm{seq}},K)\right)|\lambda]\nonumber\\
&=\sum_{k=1}^{M}k\cdot p_s(k) \left(1-p_{\textrm{coll}}(N_{\textrm{seq}},k)\right)f_K(k|\lambda).
\end{align}
Using (6), it is possible to compute the throughputs of the random access schemes considered in this paper.

\subsection{Conventional random access scheme}
With conventional random access, each remote station transmits a random access packet with constant power $P$ regardless of channel gain. Then, the received SINR of $z_0$ is computed as
\begin{equation}
SINR_{0,\textrm{conv}}=\frac{g_0}{\displaystyle\frac{N_0}{PT_\text{p}} + \frac{1}{N}\sum^{K-1}_{j=1}g_j}.
\end{equation}

 If there are $K$ simultaneous packet transmissions with conventional random access, the probability of successful packet detection $p_{s,\textrm{conv}}(K)$ is computed as
\begin{align}
p_{s,\textrm{conv}}(K)&=Pr\{SINR_{0,\textrm{conv}}>\eta_{\text{th}} |K\}\nonumber\\
 &= Pr\left\{g_0>\eta_{\text{th}}\left(\frac{N_0}{PT_\text{p}} + \frac{1}{N}\sum^{K-1}_{j=1}g_j \right) \middle|K\right\}\nonumber\\
 &= \E \left[e^{-\eta_{\text{th}}\left(\frac{N_0}{PT_\text{p}} + \frac{1}{N}\sum^{K-1}_{j=1}g_j \right)}\right]\nonumber\\
 &= e^{-\frac{\eta_{\text{th}} N_0}{PT_\text{p}}} \mathcal{L}_{W}\left(\frac{\eta_{\text{th}}}{N}\right)\nonumber\\
 &= e^{-\frac{\eta_{\text{th}} N_0}{PT_\text{p}}} \left(\frac{N}{\eta_{\text{th}} + N}\right)^{K-1},
\end{align}
where $\mathcal{L}_{W}(\cdot)$ is Laplace transform of the pdf of $W$, which is $W=\sum^{K-1}_{j=1}g_j$.

For an average arrival rate $\lambda$, throughput $S_{\textrm{conv}}(\lambda)$ of conventional random access is computed as
\begin{align}
S_{\textrm{conv}}(\lambda)&=\sum^{M}_{k=1}k p_{s,\textrm{conv}}(k) \left( 1-p_{\textrm{coll}}(N_{\textrm{seq}},k) \right) f_K(k|\lambda)\nonumber\\
&\stackrel{(a)}{\approx} \sum^{\infty}_{k=1}k e^{-\frac{\eta_{\text{th}} N_0}{PT_\text{p}}} \left(\frac{N}{\eta_{\text{th}} + N}\right)^{k-1}\left(1-\frac{1}{N_{\textrm{seq}}}\right)^{k-1} \frac{\lambda^{k}e^{-\lambda}}{k!}\nonumber\\
&= \lambda e^{-\eta_{\text{th}}\left(\frac{\lambda}{\eta_{\text{th}}+N}+\frac{N_0}{PT_\text{p}}\right)}e^{-\frac{N\lambda}{(\eta_{\text{th}}+N)N_{\textrm{seq}}}},
\end{align}
where the upper limit of summation can be replaced by infinity instead of $M$ in the approximation (a), since $kf_K(k|\lambda)$ can be neglected for $k>M$ when $M$ is much larger than $\lambda$,

\subsection{Channel-adaptive random access scheme with constant power allocation}
 With channel-adaptive random access, the pdf of equivalent small-scale channel gain $f_{G}(g|H_{1})$ is obtained as \cite{CARA}
\begin{equation}
f_{G}(g|H_1)=
  \begin{cases}
   f_G(g) + \int\limits_{0}^{g_{\text{th}}}{f_G(g)\, dg} \cdot \delta(g-g_{\text{th}}), & $if $ g\geq g_{\text{th}}, \\
   0, &$otherwise$,
  \end{cases}
\end{equation}
where $H_1$ is defined as the hypothesis when a remote station transmits a packet and $\delta(x)$ is a Dirac delta function.

 When constant power allocation is used for channel-adaptive random access, each random access packet is transmitted with constant power. Then, the received SINR of $z_0$ is computed as
\begin{equation}
SINR_{0,\textrm{const}}=\frac{g_0}{\displaystyle\frac{N_0}{P_\textrm{C}T_\text{p}} + \frac{1}{N}\sum^{K-1}_{j=1}g_j},
\end{equation}
where $P_\textrm{C}$ is the transmission power for constant power allocation.

Under the condition that $K$ remote stations transmit packets simultaneously, $W$ is defined as $\sum^{K-1}_{j=1}g_j$. Denote by $V$ the number of channel gains $g_j$ which are larger than $g_{\text{th}}$. The pdf of $W$ can be computed as \cite{Probability_berts}
  \begin{equation}
f_{W}(w|K)=\sum^{K-1}_{v=0}f_{W}(w|V=v)Pr(V=v).
\end{equation}
 Let us define $q_j=g_j-g_{\text{th}}$. Then, $q_j$ follows an exponential distribution \cite{Probability_berts}. If $v$ channel gains are larger than $g_{\text{th}}$ (that is $V=v$), $\sum^{K-1}_{j=1}q_j$ follows an Erlang distribution of order $v$ \cite{Probability_berts}. Therefore, from $w=\sum^{K-1}_{j=1}q_j+(K-1)g_{\text{th}}$, the pdf of $w$ is obtained as
  \begin{equation}
  \begin{split}
&f_{W}(w|K) \\
&=  \begin{cases}
  \sum\limits^{K-1}_{v=1} \binom{K-1}{v} \frac{\left[w-(K-1)g_{\text{th}}\right]^{v-1}e^{-\left[w-(K-1)g_{\text{th}}\right]}}{(v-1)!}\cdot e^{-vg_{\text{th}}}(1-e^{-g_{\text{th}}})^{K-1-v} \\ + (1-e^{-g_{\text{th}}})^{K-1}\delta(w-(K-1)g_{\text{th}}), & $if $ w\geq (K-1)g_{\text{th}}, \\
   0, &$otherwise$.
  \end{cases}
  \end{split}
\end{equation}

\begin{figure*}[t!]
    \centering
    \begin{subfigure}[t]{0.5\textwidth}
        \centering
         \epsfig{file=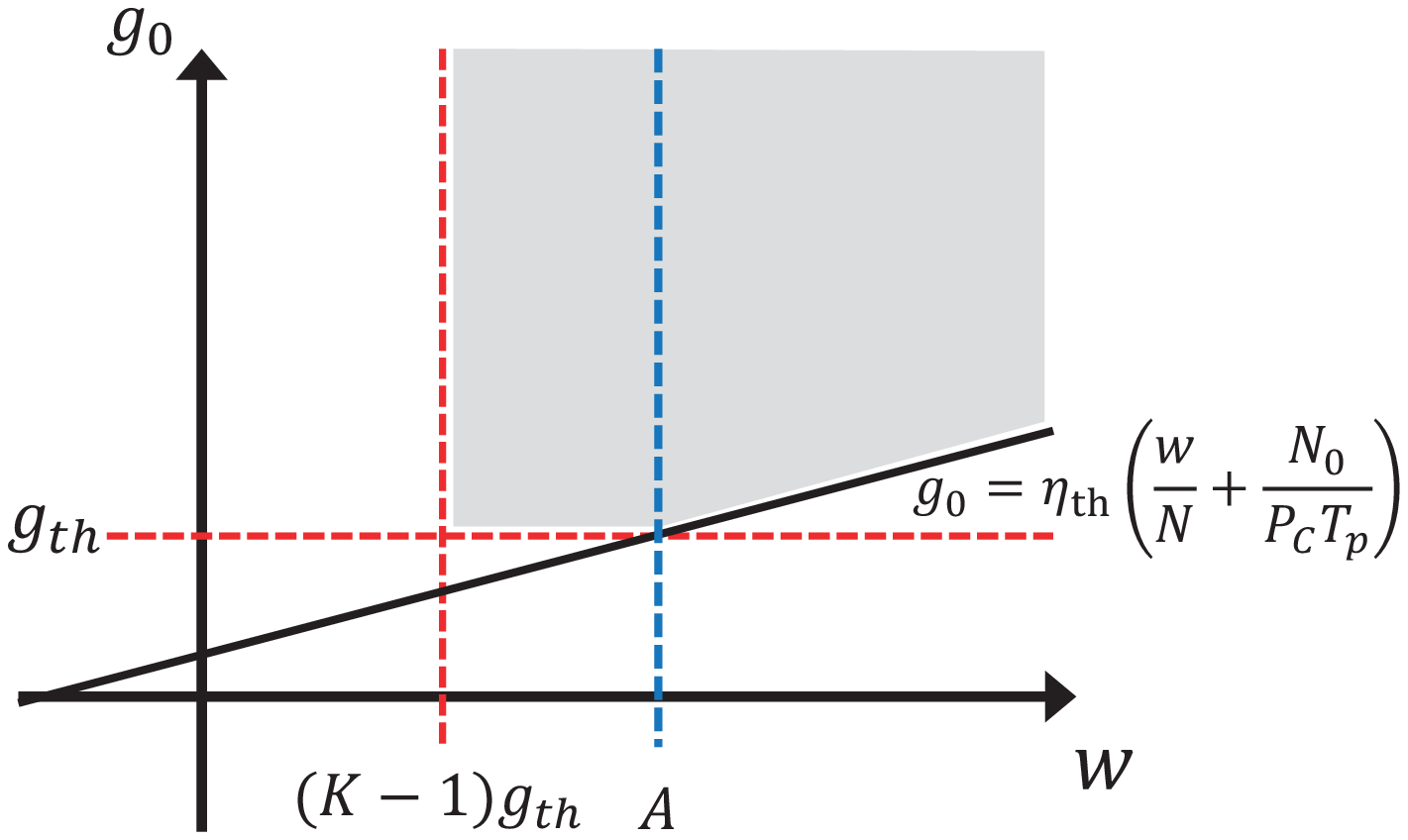, trim=0.0cm 0.0cm 0.0cm 0.0cm, clip=true, width=8.5cm}
        \caption{$K\leq 1+\frac{N}{g_{\text{th}}}\left(\frac{g_{\text{th}}}{\eta_{\text{th}}}-\frac{N_0}{P_CT_\text{p}}\right)$}
    \end{subfigure}\\
    ~
    \begin{subfigure}[t]{0.5\textwidth}
        \centering
         \epsfig{file=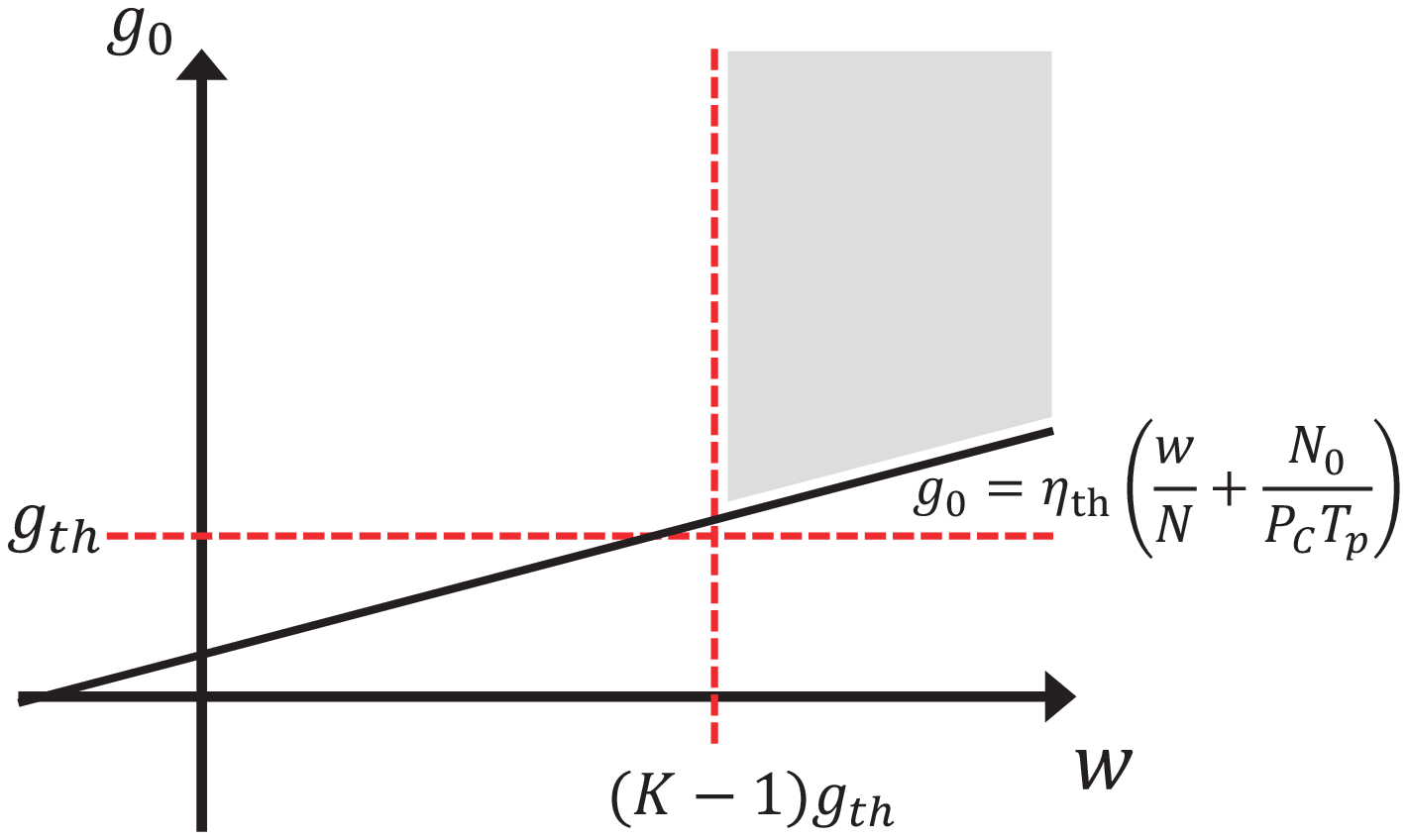, trim=0.0cm 0.0cm 0.0cm 0.0cm, clip=true, width=8.5cm}
        \caption{$K>1+\frac{N}{g_{\text{th}}}\left(\frac{g_{\text{th}}}{\eta_{\text{th}}}-\frac{N_0}{P_CT_\text{p}}\right)$}
    \end{subfigure}
    \caption{Two different cases for $g_0\geq \eta_{\text{th}} \left( \frac{N_0}{P_CT_\text{p}} +\frac{w}{N} \right)$ in $g$-$w$ planes.}
\end{figure*}

If there are $K$ simultaneous packet transmissions with channel-adaptive random access using constant power allocation, the probability of successful packet detection $p_{s,\textrm{const}}(K)$ is obtained as
\begin{equation}
\begin{split}
p_{s,\textrm{const}}(K)& =Pr\left\{ g_0\geq \eta_{\text{th}} \left( \frac{N_0}{P_CT_\text{p}} +\frac{w}{N} \right)\middle|K\right\}.
\end{split}
\end{equation}

To compute $p_{s,\textrm{const}}(K)$, it is necessary to consider two different cases: $K\leq 1+\frac{N}{g_{\text{th}}}\left(\frac{g_{\text{th}}}{\eta_{\text{th}}}-\frac{N_0}{P_CT_\text{p}}\right)$ and $K>1+\frac{N}{g_{\text{th}}}\left(\frac{g_{\text{th}}}{\eta_{\text{th}}}-\frac{N_0}{P_CT_\text{p}}\right)$, since the range of the random variable $w$ is dependent on $K$ for fixed $g_{\text{th}}$, $\eta_{\text{th}}$ and $N$ values.
Fig 2. shows the regions satisfying $g_0\geq \eta_{\text{th}} \left( \frac{N_0}{P_CT_\text{p}} +\frac{w}{N} \right)$ in $g\text{-}w$ planes for the two different cases of $K$.

Let us consider the first case when $K\leq 1+\frac{N}{g_{\text{th}}}\left(\frac{g_{\text{th}}}{\eta_{\text{th}}}-\frac{N_0}{P_CT_\text{p}}\right)$. For this case, $p_{s,\textrm{const}}(K)$ is computed as
\begin{align}
&p_{s,\textrm{const}}(K)\nonumber\\
&=\int^{A}_{(K-1)g_{\text{th}}}\int^{\infty}_{g_{\text{th}}}f_W(w)f_G(g|H_1)dgdw +\int^{\infty}_{A}\int^{\infty}_{\eta_{\text{th}}\left(\frac{w}{N}+\frac{N_0}{P_CT_\text{p}}\right)}f_W(w)f_G(g|H_1)dgdw\nonumber\\
&=\sum^{K-1}_{v=1} \binom{K-1}{v}\frac{e^{-vg_{\text{th}}}(1-e^{-g_{\text{th}}})^{K-1-v}}{(v-1)!}
 \left\{\int^{A}_{(K-1)g_{\text{th}}} \left[w-(K-1)g_{\text{th}}\right]^{v-1}e^{-\left[w-(K-1)g_{\text{th}}\right]}dw\right. \nonumber\\
& \left. +\int^{\infty}_{A}e^{-\eta_{\text{th}}\left( \frac{w}{N}+\frac{N_0}{P_CT_\text{p}}\right)} \left[w-(K-1)g_{\text{th}}\right]^{v-1}e^{-\left[w-(K-1)g_{\text{th}}\right]}dw\right\} +(1-e^{-g_{\text{th}}})^{K-1} \nonumber\\
 &=\sum^{K-1}_{v=1}\binom{K-1}{v} \frac{e^{-vg_{\text{th}}}(1-e^{-g_{\text{th}}})^{K-1-v}}{(v-1)!}\left\{ \gamma(v,B)+e^{-\eta_{\text{th}}\left( \frac{(K-1)g_{\text{th}}}{N}+\frac{N_0}{P_CT_\text{p}}\right)} \left(\frac{1}{\beta}\right)^v \Gamma(v,B\beta) \right\}\nonumber\\
 &+(1-e^{-g_{\text{th}}})^{K-1},
\end{align}
where $A=N\left( \frac{g_{\text{th}}}{\eta_{\text{th}}}-\frac{N_0}{P_CT_\text{p}}\right)$, $\beta=1+\frac{\eta_{\text{th}}}{N}$, $B=A-(K-1)g_{\text{th}}$,  $\gamma(s,x) = \int_0^x t^{s-1}\,e^{-t} dt$ and $\Gamma(s,x) = \int_x^{\infty} t^{s-1}\,e^{-t}dt$ \cite{incomGamma}.

Next, consider the other case when $K>1+\frac{N}{g_{\text{th}}}\left(\frac{g_{\text{th}}}{\eta_{\text{th}}}-\frac{N_0}{P_CT_\text{p}}\right)$. Then, $p_{s,\textrm{const}}(K)$ is computed as
\begin{align}
&p_{s,\textrm{const}}(K)\nonumber\\
&=\int^{\infty}_{(K-1)g_{\text{th}}}\int^{\infty}_{\eta_{\text{th}}\left(\frac{w}{N}+\frac{N_0}{P_CT_\text{p}}\right)}f_W(w)f_G(g|H_1)dgdw\nonumber\\
&= \sum^{K-1}_{v=1} \binom{K-1}{v}\frac{e^{-vg_{\text{th}}}(1-e^{-g_{\text{th}}})^{K-1-v}}{(v-1)!}\cdot e^{-\eta_{\text{th}}\left( \frac{(K-1)g_\textrm{th}}{N}+\frac{N_0}{P_CT_\text{p}}\right)}\int^{\infty}_{0} u^{v-1}e^{-\left(1+\frac{\eta_{\text{th}}}{N}\right)u}du \nonumber\\
&= e^{-\eta_{\text{th}}\left( \frac{(K-1)g_{\text{th}}}{N}+\frac{N_0}{P_CT_\text{p}}\right)}  \left(1-e^{-g_{\text{th}}}\frac{\eta_{\text{th}}}{N+\eta_{\text{th}}}\right)^{K-1},
\end{align}
where $u=w-(K-1)g_{\text{th}}$.

Therefore, the probability of successful packet detection $p_{s,\textrm{const}}(K)$ is expressed as
\begin{equation}
\begin{split}
&p_{s,\textrm{const}}(K)\\
&= \begin{cases}
  (1-e^{-g_{\text{th}}})^{K-1} + \sum\limits^{K-1}_{v=1}\binom{K-1}{v} \frac{e^{-vg_{\text{th}}}(1-e^{-g_{\text{th}}})^{K-1-v}}{(v-1)!}\\ \cdot\left\{ \gamma(v,B)+e^{-\eta_{\text{th}}\left( \frac{(K-1)g_{\text{th}}}{N}+\frac{N_0}{P_CT_{p}}\right)} \left(\frac{1}{\beta}\right)^v \Gamma(v,B\beta) \right\}, & K\leq 1+\frac{N}{g_{\text{th}}}\left(\frac{g_{\text{th}}}{\eta_{\text{th}}}-\frac{N_0}{P_CT_\text{p}}\right), \\
  e^{-\eta_{\text{th}}\left( \frac{(K-1)g_{\text{th}}}{N}+\frac{N_0}{P_CT_{p}}\right)} \cdot \left(1-e^{-g_{\text{th}}}\frac{\eta_{\text{th}}}{N+\eta_{\text{th}}}\right)^{K-1}, &K>1+\frac{N}{g_{\text{th}}}\left(\frac{g_{\text{th}}}{\eta_{\text{th}}}-\frac{N_0}{P_CT_\text{p}}\right).
\end{cases}
\end{split}
\end{equation}

Throughput $S_{\textrm{const}}(\lambda)$ can be computed by substituting (2) and (19) into (8) for channel-adaptive random access using constant power allocation.

\subsection{Channel-adaptive random access scheme with channel inversion power allocation}
When channel inversion power allocation is used for channel-adaptive random access, the received power of each random access packet becomes a constant $P_I$ at a host station. Therefore, in this case, the received power for a packet is independent of channel gain. Therefore, the received SINR of $z_0$ is computed as
 \begin{equation}
 SINR_{0,\textrm{inv}} = \frac{1}{\displaystyle\frac{N_0}{P_IT_{p}} + \frac{K-1}{N}}.
 \end{equation}

If there are $K$ simultaneous packet transmissions, the probability of successful packet detection $p_{s,\textrm{inv}}(K)$ is computed as
\begin{equation}
\begin{split}
p_{s,\textrm{inv}}(K)&=Pr\{ SINR_{0,\textrm{inv}}>\eta_{\text{th}} |K\}\\&=
  \begin{cases}
   1,   \,\,\,\,\, & \text{for    } K< \floor[\Big]{1+N \left(\frac{1}{\eta_{\text{th}}}-\frac{N_0}{P_IT_\text{p}}\right)}, \\
   0,     &  \text{for    } K\geq \floor[\Big]{1+N \left(\frac{1}{\eta_{\text{th}}}-\frac{N_0}{P_IT_{p}}\right)},
  \end{cases}
  \end{split}
 \end{equation}
 where $\floor{ x }$ is the largest integer that is not larger than $x$.

For a fixed average arrival rate $\lambda$, throughput $S_{\textrm{inv}}(\lambda)$ is computed as
\begin{align}
S_{\textrm{inv}}(\lambda)&=\sum^{\infty}_{k=1}k p_{s,\textrm{inv}}(k) \left( 1-p_{\textrm{coll}}(N_{\textrm{seq}},k) \right) f_K(k|\lambda)\nonumber\\
&=\sum^{\floor[\Big]{1+N \left(\frac{1}{\eta_{\text{th}}}-\frac{N_0}{P_IT_{p}}\right)}}_{k=1}k \left(1-\frac{1}{N_{\textrm{seq}}}\right)^{k-1} \frac{\lambda^{k}e^{-\lambda}}{k!}\nonumber\\
&=C\lambda e^{-\lambda},
\end{align}
where $C=\sum^{\floor[\Big]{1+N \left(\frac{1}{\eta_{\text{th}}}-\frac{N_0}{P_IT_\text{p}}\right)}}_{k=1} \frac{\left(\lambda-\frac{\lambda}{N_{\textrm{seq}}}\right)^{k-1}}{(k-1)!}.$

As shown in (20), throughput $S_{\text{inv}}(\lambda)$ is independent of the transmission threshold $g_{\text{th}}$ for channel-adaptive random access using channel inversion power allocation.
\section{Numerical Result}
In this section, numerical results are presented for several random access schemes. The results are obtained from both analysis and simulation. Analytic results are computed from the equations derived in section \rom{3}. A simulator was built to obtain the numerical results.

Table \rom{1} summarizes the parameters used for the simulator. There are total 8192 remote stations ($M=8192$) in a system. The number of simultaneous packets $K$ follows a Poisson distribution with an average arrival rate $\lambda$ per random access slot. Each packet is transmitted over an independent Rayleigh fading channel. The processing gain $N$ of a random access packet is set to 64, and the minimum required SINR $\eta_{\text{th}}$ is assumed to be $1.0, 5.0$ or 10.0 dB for the SINR capture model. A transmission threshold $g_{\text{th}}$ is set to satisfy specific outage probability $P_\text{o}=Pr\{g<g_{\text{th}}\}$.

Throughput is dependent on both thermal noise $N_0$ and the interference from other remote stations. Generally, it is possible to achieve higher throughput, as thermal noise power decreases compared with interference power. Furthermore, most practical wireless systems are designed such that thermal noise is a minor factor for the throughput of random access. Therefore, to observe the relation between throughput and interference more clearly, numerical results are obtained for average $P(g)T_\text{p}/N_0= 30.0$ dB, which is a high SNR case.

\begin{table}[t]
\caption{Simulation parameters}
\centering
\begin{tabular}{c c}
\hline
Number of remote stations $M$     & 8192        \\
Number of simultaneous packets $K$        & Poisson distribution        \\
Fading channel       & Independent Rayleigh fading channel     \\
Processing gain $N$ & 64             \\
Minimum required SINR $\eta_{\text{th}}$  &   1.0, 5.0 and 10.0 dB\\
Outage probability $P_\text{o}$ &   0.0$\sim$1.0\\
Number of sequences $N_{\textrm{seq}}$  &  $2^x$ ($1\leq x \leq 14$)\\
Average $P(g)T_\text{p}/N_0$   &   30.0 dB\\
\hline
\end{tabular}
\end{table}

 \begin{figure}[t!]
  \centering
     \epsfig{file=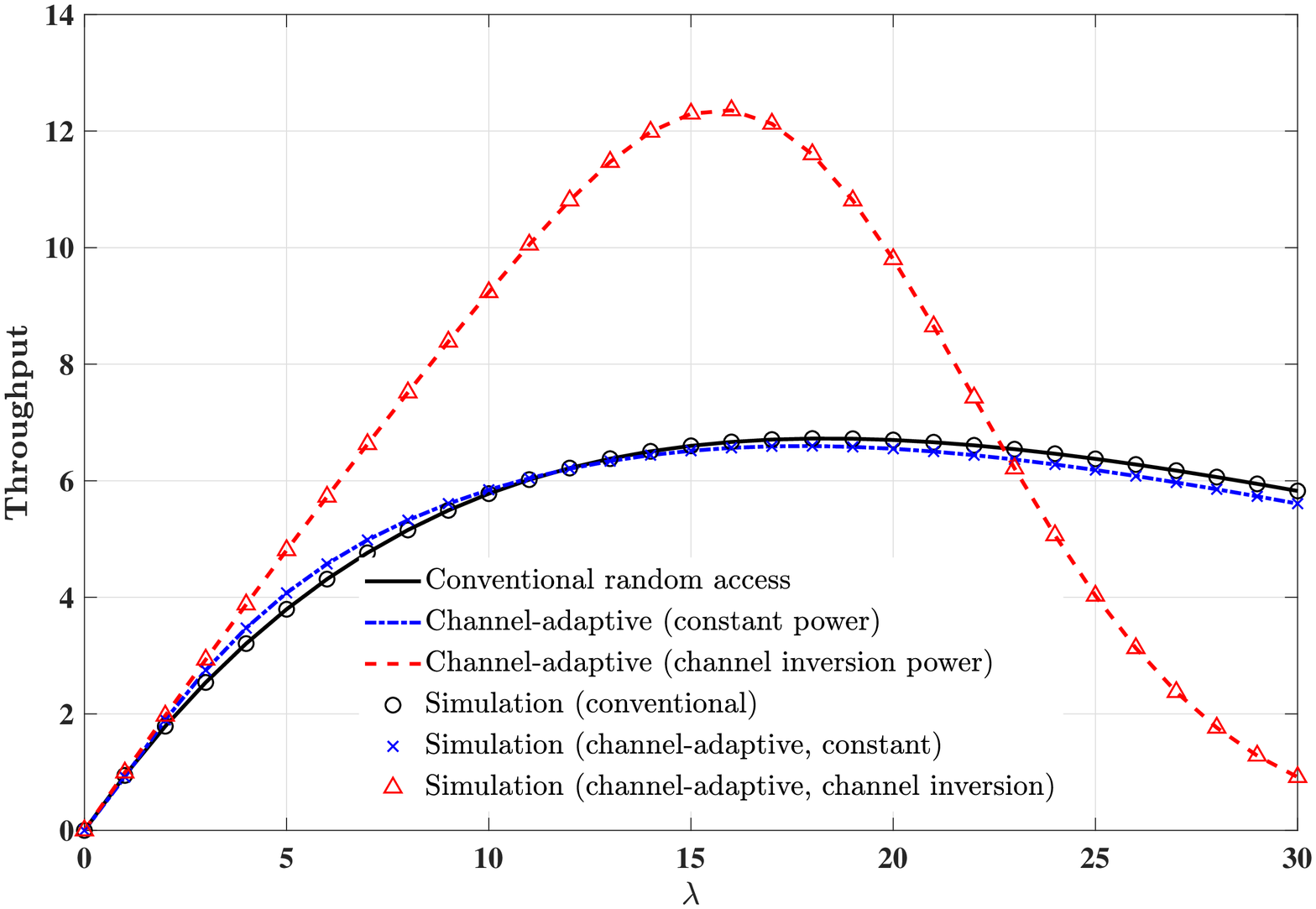, trim=0.0cm 0.0cm 0.0cm 0.0cm, clip=true, width=12cm}
	\caption{Throughput versus average arrival rate. ($P_\text{o}=0.2$, $N_{\text{seq}}=128$)}
\end{figure}
 \begin{figure}
  \centering
     \epsfig{file=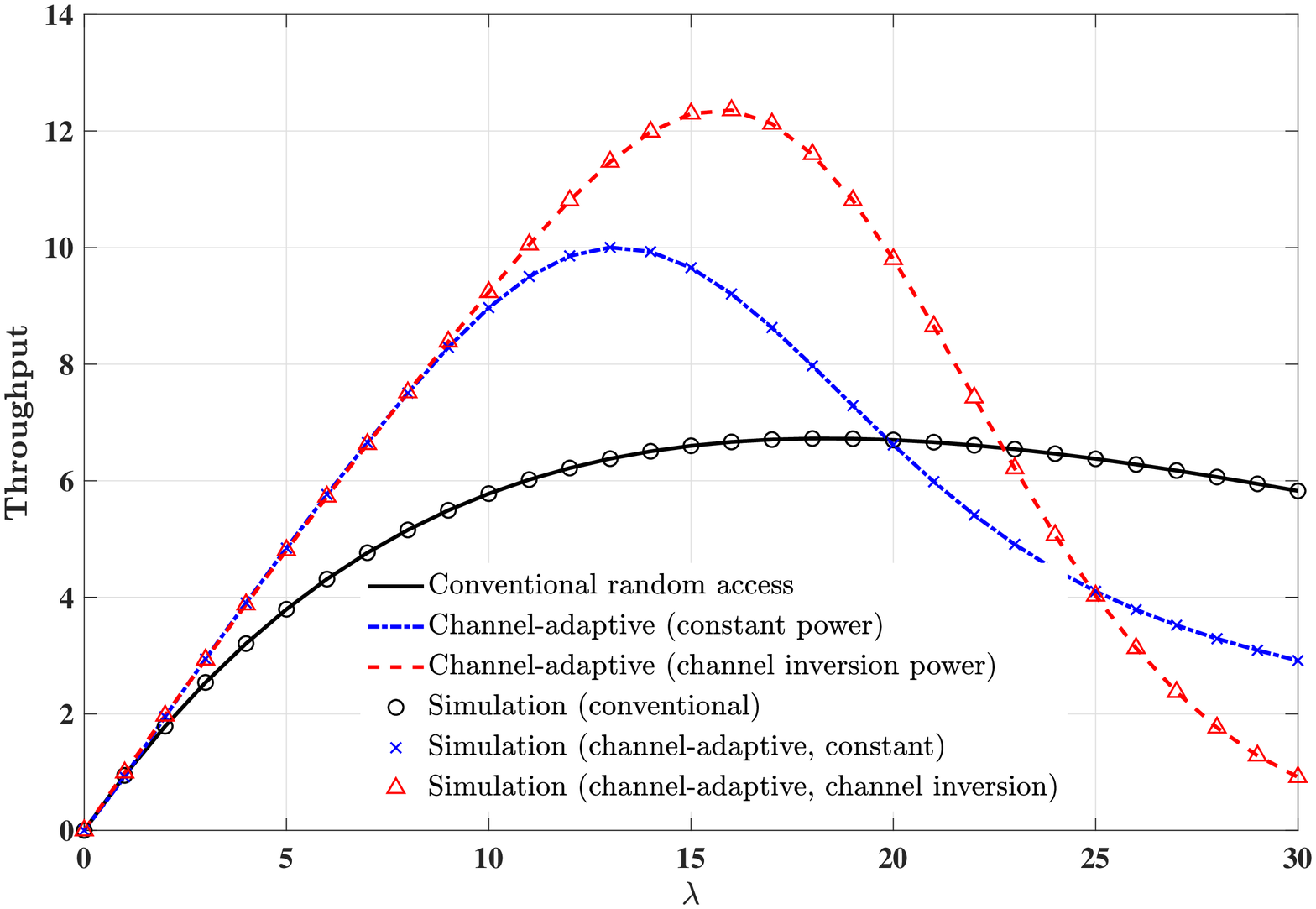, trim=0.0cm 0.0cm 0.0cm 0.0cm, clip=true, width=12cm}
	\caption{Throughput versus average arrival rate. ($P_\text{o}=0.7$, $N_{\text{seq}}=128$)}
\end{figure}
 \begin{figure}
  \centering
     \epsfig{file=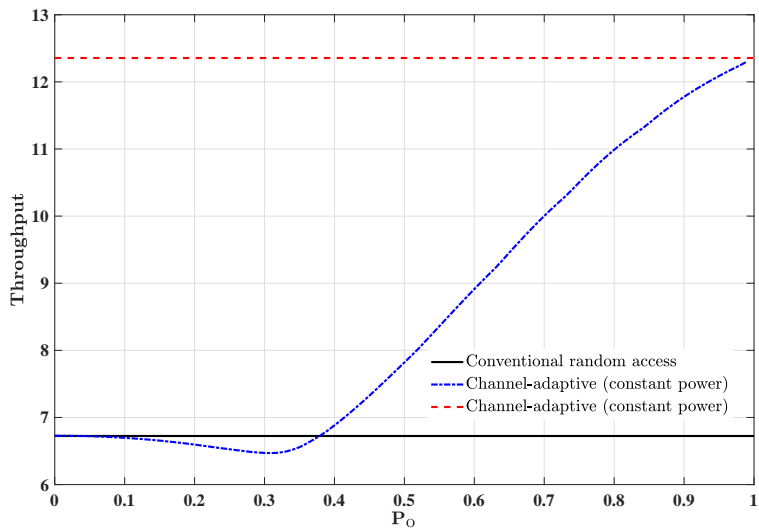, trim=0.0cm 0.0cm 0.0cm 0.0cm, clip=true, width=12cm}
	\caption{$\max_{\lambda}S(5.0, P_{\text{o}}, 128)$ versus $P_\text{o}$.}
\end{figure}
 \begin{figure}
  \centering
     \epsfig{file=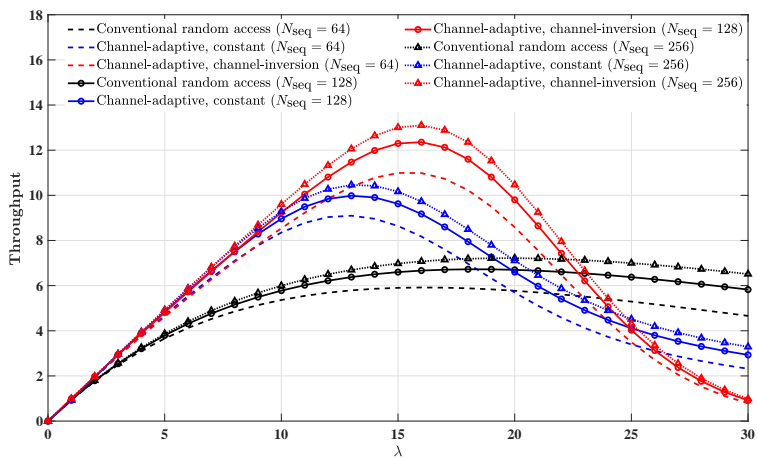, trim=0.0cm 0.0cm 0.0cm 0.0cm, clip=true, width=12cm}
	\caption{Throughput versus average arrival rate. ($P_\text{o}=0.7$)}
\end{figure}
 \begin{figure}
  \centering
     \epsfig{file=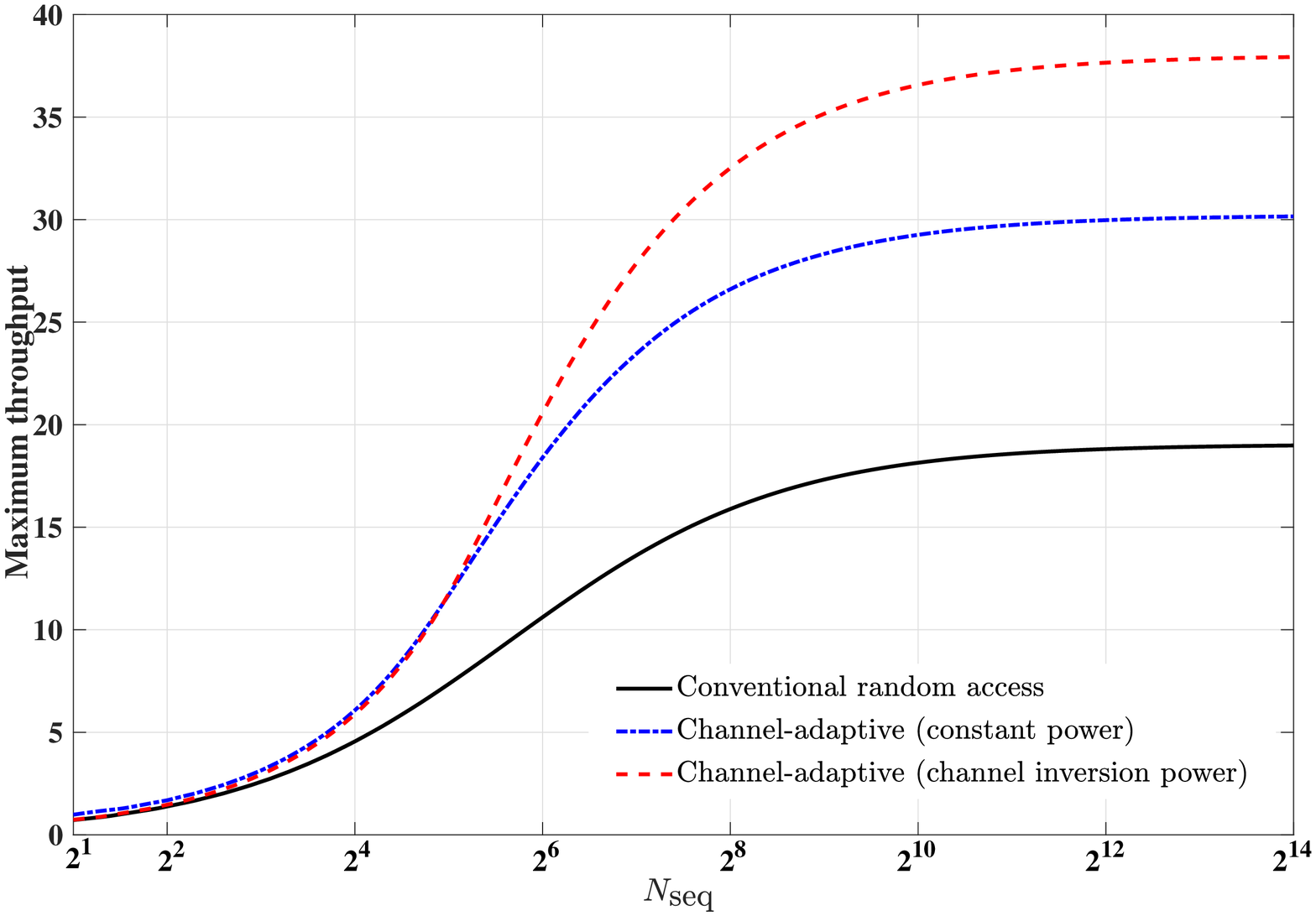, trim=0.0cm 0.0cm 0.0cm 0.0cm, clip=true, width=12cm}
	\caption{Maximum throughput versus $N_{\textrm{seq}}$. ($P_\text{o}=0.7$, $\eta_{\text{th}}= 1.0$ dB)}
\end{figure}
 \begin{figure}
  \centering
     \epsfig{file=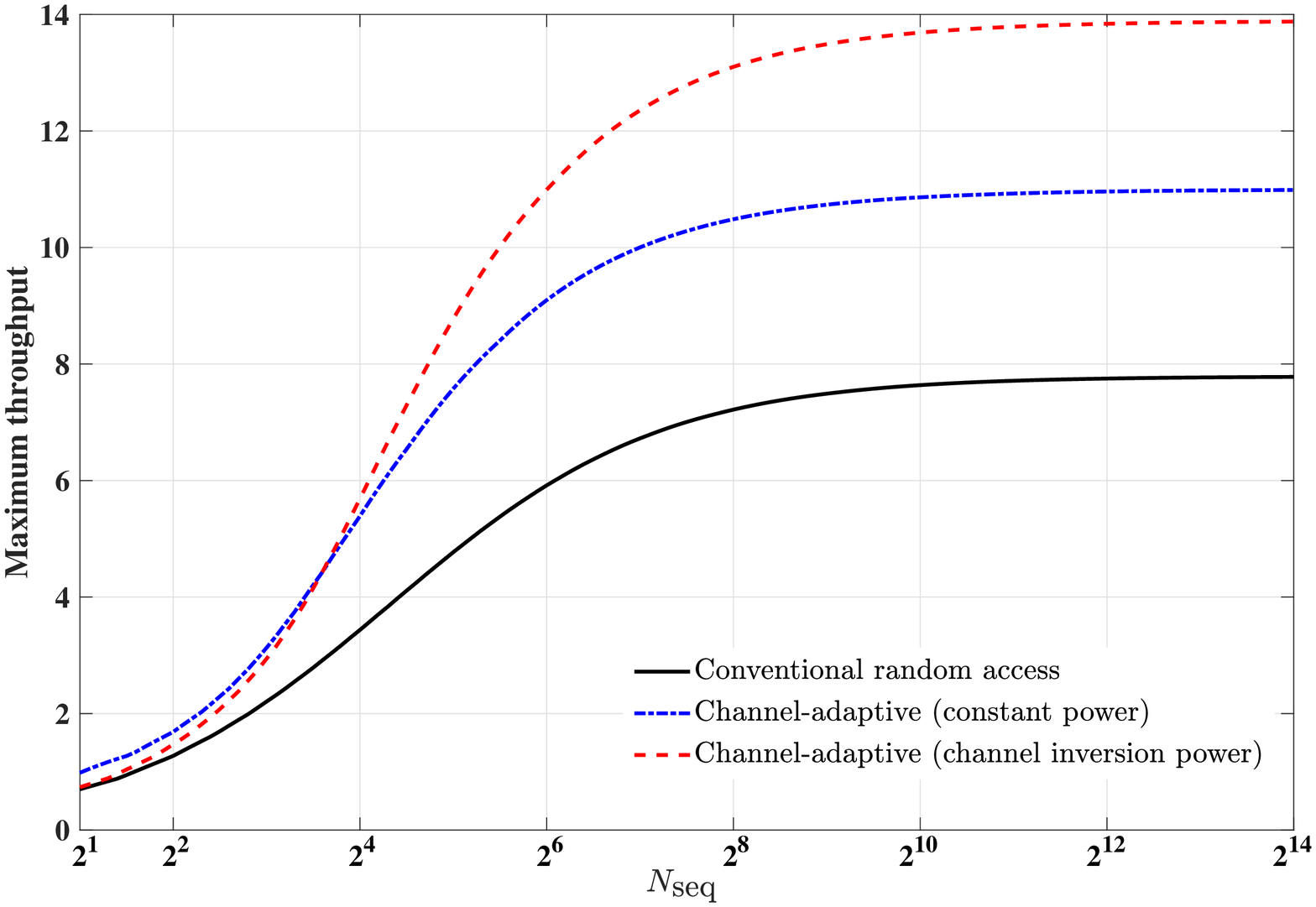, trim=0.0cm 0.0cm 0.0cm 0.0cm, clip=true, width=12cm}
	\caption{Maximum throughput versus $N_{\textrm{seq}}$. ($P_\text{o}=0.7$, $\eta_{\text{th}}= 5.0$ dB)}
\end{figure}
 \begin{figure}
  \centering
     \epsfig{file=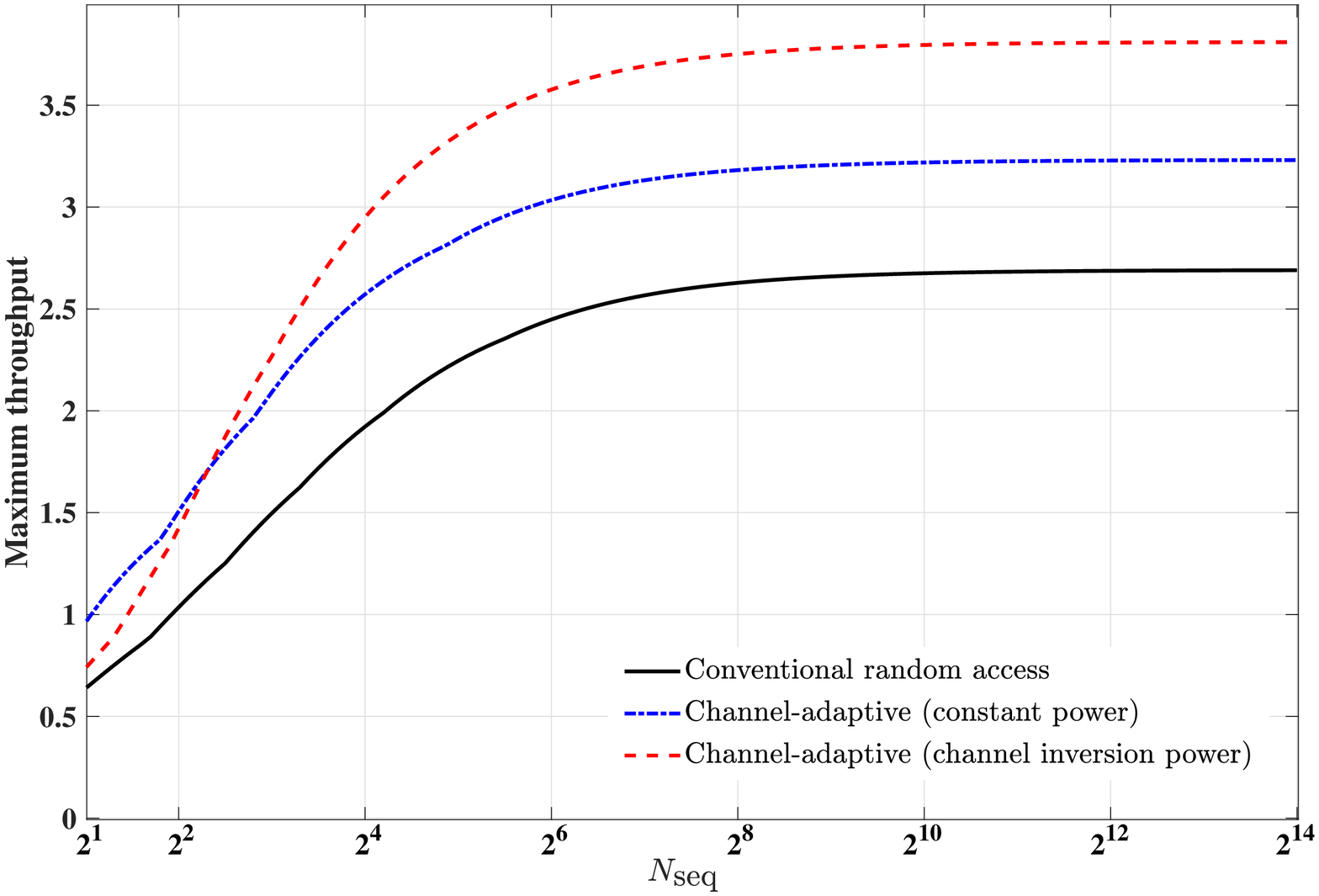, trim=0.0cm 0.0cm 0.0cm 0.0cm, clip=true, width=12cm}
	\caption{Maximum throughput versus $N_{\textrm{seq}}$. ($P_\text{o}=0.7$, $\eta_{\text{th}}= 10.0$ dB)}
\end{figure}

Figs. 3 and 4 show throughputs versus $\lambda$ for several random access schemes with $P_\text{o}=0.2$ and $0.7$, respectively. The results of Figs. 3 and 4 are obtained with the number of sequences $N_{\text{seq}}=128$ and $\eta_{\text{th}}= 5.0$ dB.

 In Fig. 3, the results are obtained for $P_{\text{o}}=0.2$. For comparison, the throughput of conventional random access is added. With conventional random access, peak throughput is 6.7 for $\lambda=18.1$. With channel-adaptive random access using constant and channel inversion power allocations, peak throughputs are 6.6 and 12.4 for $\lambda=17.9$ and $15.8$, respectively.

 In Fig. 4, the results are obtained for $P_{\text{o}}=0.7$. For comparison, the throughput of conventional random access is added. With conventional random access and channel-adaptive random access using channel inversion power allocation, throughput curves are exactly the same as those in Fig. 3. However, with channel-adaptive random access using constant power allocation, for $\lambda=12.9$, peak throughput of Fig. 4 is 10.0, which is about 1.5 times higher than that of Fig. 3. This is because interference power decreases significantly as the transmission threshold $g_{\text{th}}$ increases.

From Figs. 3 and 4, it can be observed that there is almost no difference between the simulation and analytic results. In addition, with channel-adaptive random access, peak throughput is obtained at a lower average arrival rate $\lambda$ compared to conventional random access.

Comparing Figs. 3 and 4, it is possible to observe that only the throughput of channel-adaptive random access using constant power allocation is dependent on $P_{\text{o}}$ among the schemes considered in this paper. In Fig. 3, its throughput curve is similar to that of conventional random access.
In Fig. 4, its throughput curve becomes closer to that of channel inversion power allocation.
As the $P_{\text{o}}$ value approaches 1.0, with channel-adaptive random access, it is noticed that the difference in throughput decreases between constant and channel inversion power allocations.

Let us consider the throughput of channel-adaptive random access using channel inversion power allocation. In Figs. 3 and 4, its throughput is proportional to $\lambda$ for $\lambda < 10.0$. After reaching a peak, its throughput decreases more rapidly than the other schemes. For $\lambda$ larger than 22.6, throughput is lower than that of conventional random access. This is because, with a SINR capture model, the probability of successful packet detection becomes zero from (21), when the number of simultaneous packet transmissions $K$ is larger than a certain value.

Throughput depends on several factors including an average arrival rate $\lambda$, a minimum required SINR $\eta_{\text{th}}$, outage probability $P_{\text{o}}$ and the number of sequences $N_{\text{seq}}$.
Let us denote by $\max_{\lambda}S(\eta_{\text{th}},P_{\text{o}}, N_{\text{seq}})$ the maximum throughput over $\lambda$ under given $\eta_{\text{th}}$, $N_{\text{seq}}$ and $P_{\text{o}}$ values.

  Fig. 5 shows $\max_{\lambda}S(\eta_{\text{th}},P_{\text{o}}, N_{\text{seq}})$ versus outage probability $P_\text{o}$ for several random access schemes with fixed $\eta_{\text{th}}= 5.0$ dB and $N_{\text{seq}}=128$. With conventional random access, $\max_{\lambda}S(5.0, P_{\text{o}}, 128)$ is 6.7. Channel-adaptive random access using channel inversion power allocation achieves about 1.85 times higher $\max_{\lambda}S(5.0, P_{\text{o}}, 128)$ than conventional random access.

  In Fig. 5, it can be also observed that throughput is independent of $P_\text{o}$ with channel-adaptive random access using channel inversion power allocation. However, with channel-adaptive random access using constant power allocation, $\max_{\lambda}S(5.0, P_{\text{o}}, 128)$ is highly dependent on $P_\text{o}$. This value decreases to 6.5 for $P_\text{o} \in [0,0.31]$, and then increases to 12.4 for $P_\text{o} \in (0.31,1.0]$. For $P_\text{o} \in [0,0.38]$, it is slightly lower than that of conventional random access. However, except that range, this value is larger than that of conventional random access.

Fig. 6 shows the throughputs of several random access schemes for different numbers of sequences $N_{\text{seq}}$ ($N_{\text{seq}}=$64, 128 or 256). From Fig. 6, it can be observed that throughput is dependent on the number of sequences. For the results, $P_\text{o}$ and $\eta_{\text{th}}$ are set to 0.7 and 5.0 dB, respectively. In Fig. 6, with conventional random access, the maximum throughputs are 5.9, 6.7 and 7.2, for $N_{\textrm{seq}}=$64, 128 and 256, respectively. When constant power allocation is used, channel-adaptive random access achieves about 1.56, 1.49 and 1.45 times higher maximum throughput than conventional random access, for $N_{\textrm{seq}}=$64, 128 and 256, respectively. When channel inversion power allocation is used, channel-adaptive random access achieves about 1.86, 1.85 and 1.82 times lager maximum throughput than conventional random access, for $N_{\textrm{seq}}=$64, 128 and 256, respectively. As the number of sequences $N_{\text{seq}}$ increases, throughput increases for all random access schemes, since the probability of sequence collision decreases.

 As the number of sequences $N_{\text{seq}}$ increases to infinity, $\max_{\lambda}S(\eta_{\text{th}},P_{\text{o}}, N_{\text{seq}})$ converges to a certain value for fixed $\eta_{\text{th}}$ and $P_{\text{o}}$ values. This is the achievable maximum throughput, when there is no sequence collision. Let us denote $T_{\text{conv}}$, $T_{\text{const}}$ and $T_{\text{inv}}$ as $\lim_{N_{\text{seq}} \to \infty}$ $\max_{\lambda}S(\eta_{\text{th}}, P_{\text{o}}, N_{\textrm{seq}})$ with conventional random access and channel-adaptive random access using constant and channel inversion power allocations, respectively.

Figs. 7, 8 and 9 show $\max_{\lambda}S(\eta_{\text{th}}, P_{\text{o}}, N_{\textrm{seq}})$ versus $N_{\textrm{seq}}$ for several random access schemes with different minimum required SINR $\eta_{\text{th}}$ values. With these figures, it is possible to observe the dependency of throughput on the minimum required SINR $\eta_{\text{th}}$ and the number of sequences $N_{\text{seq}}$. For the results, $P_\text{o}$ is fixed to 0.7 and $\eta_{\text{th}}$ is set to 1.0 5.0 or 10.0 dB.

Fig. 7 shows $\max_{\lambda}S(\eta_{\text{th}}, P_{\text{o}}, N_{\textrm{seq}})$ versus $N_{\textrm{seq}}$ when $\eta_{\text{th}}$ is set to 1.0 dB. $T_{\text{conv}}$, $T_{\text{const}}$ and $T_{\text{inv}}$ are 19.0, 30.2 and 37.9, respectively. It is shown that about 128 sequences are necessary to achieve 80$\%$ of $T_{\text{conv}}$, $T_{\text{const}}$ and $T_{\text{inv}}$.

Fig. 8 shows $\max_{\lambda}S(\eta_{\text{th}}, P_{\text{o}}, N_{\textrm{seq}})$ versus $N_{\textrm{seq}}$ when $\eta_{\text{th}}$ is set to 5.0 dB. $T_{\text{conv}}$, $T_{\text{const}}$ and $T_{\text{inv}}$ are 7.8, 11.0 and 13.9, respectively. It is shown that about 64 sequences are necessary to achieve 80$\%$ of $T_{\text{conv}}$, $T_{\text{const}}$ and $T_{\text{inv}}$.

Fig. 9 shows $\max_{\lambda}S(\eta_{\text{th}}, P_{\text{o}}, N_{\textrm{seq}})$ versus $N_{\textrm{seq}}$ when $\eta_{\text{th}}$ is set to 10.0 dB. $T_{\text{conv}}$, $T_{\text{const}}$ and $T_{\text{inv}}$ are 2.69, 3.23 and 3.81, respectively. It is shown that about 20 sequences are necessary to achieve 80$\%$ of $T_{\text{conv}}$, $T_{\text{const}}$ and $T_{\text{inv}}$.

From Figs. 7, 8 and 9, it can be obtained that channel-adaptive random access can achieve higher $\max_{\lambda}S(\eta_{\text{th}},P_{\text{o}}, N_{\text{seq}})$ than conventional random access. Also, it is observed that throughput decreases, as a $\eta_{\text{th}}$ value increases.
When $N_{\text{seq}}$ is small, sequence collision is a more dominant factor for throughput than interference. Therefore, similar throughput is observed between conventional and channel-adaptive random access. This is because channel-adaptive random access reduces the interference from other packets, but does not reduce sequence collision. On the other hand, as $N_{\text{seq}}$ increases, interference becomes the most dominant factor for throughput. Hence, the difference in throughputs increases between conventional and channel-adaptive random access. It can be observed that throughput converges to a constant value (which is $T_{\text{conv}}$, $T_{\text{const}}$ or $T_{\text{inv}}$), as $N_{\text{seq}}$ increases to infinity. This is the case when throughput is dependent only on interference since there is almost no sequence collision.
Furthermore, as $\eta_{\text{th}}$ increases, it can be observed that the random access schemes require less number of sequences to achieve a certain fraction of $T_{\text{conv}}$, $T_{\text{const}}$ and $T_{\text{inv}}$. For example, to achieve 80\% of $T_{\text{conv}}$, $T_{\text{const}}$ and $T_{\text{inv}}$, about 128, 64 and 20 sequences are required with $\eta_{\text{th}}=$1.0, 5.0 and 10.0 dB, respectively.

\section{Conclusion}
In this paper, throughput is analyzed for several CDM-based random access schemes considering both sequence collision and interference. It is assumed that when multiple remote stations transmit random access packets with the same sequence, none of the packets are successfully detected due to sequence collision. When there is no sequence collision, an SINR capture model is used to calculate probability of successful packet detection. The probability of successful packet detection is derived for conventional random access and channel-adaptive random access. Based on probabilistic analysis, throughput is computed for conventional random access and channel-adaptive random access. The results show that channel-adaptive random access can achieve higher throughput compared to conventional random access. Especially, when channel inversion power allocation is used, throughput of channel-adaptive random access is about 1.85 times higher than that of conventional random access. Throughput of random access is dependent on several parameters including an average packet arrival rate, a required SINR, a transmission threshold and the number of sequences. To observe the characteristics of random access throughput, results are obtained with different sets of parameters. The results show the dependency of throughput on sequence collision and interference for different number of sequences. From the results, it is possible to estimate a reasonable number of sequences considering the trade-off between implementation complexity and throughput.

\end{document}